\documentclass[a4paper]{article}

\begin {document}

\title {\bf The Three-Dimensional Quantum Hamilton-Jacobi 
Equation and Microstates }
\author{A.~Bouda\footnote{Electronic address: 
{\tt bouda\_a@yahoo.fr}} 
\ and A.~ Mohamed Meziane\footnote{Electronic address: 
{\tt amohamed\_meziane@yahoo.fr}}\\
Laboratoire de Physique Th\'eorique, Universit\'e de B\'eja\"\i a,\\ 
Route Targa Ouazemour, 06000 B\'eja\"\i a, Algeria\\}

\date{\today}

\maketitle
  
\begin{abstract}
\noindent 
In a stationary case and for any potential, we solve the 
three-dimensional quantum Hamilton-Jacobi equation in terms 
of the solutions of the corresponding Schr\"odinger equation. 
Then, in the case of separated variables, by requiring that the 
conjugate momentum be invariant under any linear 
transformation of the solutions of the Schr\"odinger equation 
used in the reduced action, we clearly identify the integration 
constants successively in one, two and three dimensions. In 
each of these cases, we analytically establish that the quantum 
Hamilton-Jacobi equation describes microstates not detected 
by the Schr\"odinger equation in the real wave function case.

\end{abstract}

\vskip\baselineskip



\newpage

\section{Introduction}

Microstates were first introduced by Floyd \cite{fl1a,fl1b,fl2a,fl2b} 
and investigated by other authors \cite{fm1a,fm1b,ca,b1}. They represent 
physical states predicted by the quantum Hamilton-Jacobi equation 
but not detected by the Schr\"odinger wave function. Up to now, 
all the analytical descriptions of microstates are considered in one 
dimension. In this paper, one of our principal objective is their 
description in higher dimensions.  

Recently, quantum mechanics was  derived from an equivalence 
postulate in the one-dimensional space by Faraggi-Matone 
\cite{fm1a,fm1b,fm2a,fm2b,fm2c,fm2d,fm2e}. 
These authors, together with Bertoldi, extended 
their finding to higher dimensions \cite{bfm}. 
In particular, they established a new version of the quantum 
stationary Hamilton-Jacobi equation (QSHJE) given by the 
two relations
\begin {equation}
 {1\over 2m}\left(\vec\nabla S_{0}\right)^2-
{\hbar^2\over 2m}{\Delta R\over R}+V(x,y,z)=E,
\end {equation}
\begin {equation}
\vec {\nabla}\cdot(R^2\vec{\nabla}S_{0})=0, \hskip96pt
\end {equation}
for a non-relativistic spinless particle of mass $m$ and energy 
$E$ in an external potential $V(x,y,z)$. Relations (1) and (2) 
represent a new version of the QSHJE because the reduced action 
$S_0$ and the function $R$ are related to the Schr\"odinger wave 
function by
\begin{equation}
\Psi =R\left[\alpha \ \exp\left(i{{S_0}\over{\hbar}}\right)
+\beta \ \exp\left(-i{{S_0}\over{\hbar}}\right)\right],
\end{equation}
$\alpha$ and $\beta$ being complex constants. This relation 
is also reproduced in ref. \cite{b1} where (1) and (2) are 
derived from the Schr\"odinger equation (SE) by appealing to the 
probability current. By setting
\begin{equation}
\alpha=|\alpha| \ \exp(ia), \hskip 30pt  \beta=|\beta| \ \exp(ib),
\end{equation}
where $a$ and $b$ are real parameters, expression (3) of the 
wave function can be written as \cite{b1}
\begin {eqnarray}
\Psi =R \ \exp{\left(i{a+b\over2}\right)}\left[\left(|\alpha|
+|\beta|\right)\cos\left({S_0\over\hbar}+{a-b\over 2}\right) \right.
\hskip20mm&& \nonumber \\
\left. +i\left(|\alpha|-|\beta|\right)\sin\left({S_0\over\hbar}
+{a-b\over 2}\right)\right].
\end {eqnarray}
In Bohm's theory \cite{bohm1,bohm2}, in which $\alpha=1$ and $\beta=0$, 
the reduced action $S_0$ is a constant in the case where the 
wave function is real, up to a constant phase factor. However, 
with expression (3), $S_0$ is never constant. In particular,  
we clearly see from (5) that the reality of the wave function  
is expressed by $|\alpha|=|\beta|$ and not by $S_0=cte$.

On the other hand, many suggestions to formulate the quantum 
trajectory equations were proposed \cite{fl2a,fl2b,bohm1,bd1a,bd1b,bd2,bh}. 
In a recent paper \cite{b2}, the QSHJE in one dimension,
\begin{eqnarray}
{1\over2m}{\left(\partial{S_0}\over\partial{x}\right)^2}+
V\left(x\right)-E=\hskip60mm&& \nonumber \\
{\hbar ^2\over4m}\left[{{3\over2}
\left(\partial{S_0}\over{\partial{x}}\right)^{-2}
\left(\partial^2{S_0}\over\partial{x^2}\right)^2}-
{\left(\partial{S_0}\over\partial{x}\right)^{-1}}
{\left(\partial^3{S_0}\over\partial{x^3}\right)}\right],
\end{eqnarray}
is reproduced from a general lagrangian depending on  
coordinate $x$ and its higher temporal derivatives 
$(\dot{x},\ddot{x},...)$ by appealing to the dimensional 
analysis. In particular, it is analytically established 
that the resulting quantum law of motion is
\begin{equation}
m\dot{x}={\partial{S_{0}}\over\partial{x}},
\end{equation}
recalling the Bohm relation. However, in contrast to Bohm's 
theory \cite{bohm1,bohm2} where $S_0$ is deduced from the wave function (3) by 
setting $\alpha=1$ and $\beta=0$, in ref. \cite{b2}, $S_0$ 
represents the solution of the third order differential equation 
$(6)$. The extension of relation (7) to three dimensions can be 
sensibly assumed as
\begin{equation}
m\dot{x}={\partial{S_{0}}\over\partial{x}},\hskip20pt m\dot{y}=
{\partial{S_{0}}\over\partial{y}},\hskip20pt m\dot{z}=
{\partial{S_{0}}\over\partial{z}}.
\end{equation}
Here, $S_0$ must be a solution of the 
couple of relations (1) and (2). 

The paper is organized as follows. In section 2, we solve 
in three dimensions the QSHJE for any potential. In section 3, 
we consider the case of separated variables and identify the 
integration constants of the reduced action. We then investigate 
in section 4 microstates in one and higher dimensions. 
Finally, we devote section 5 to conclusion.

\section{ The three-dimensional solution of the QSHJE}

From eq. (5), we can deduce that \cite{b1}
\begin{equation}
{S_0}={\hbar\hskip 2pt \arctan}{\left({{|\alpha|+
|\beta|}\over{|\alpha|-|\beta|}}
\hskip 4pt{{{\rm \ Im} \ \left[\exp\left(-i(a+b)/2\right) \ \Psi\right]}
\over{{\rm \ Re} \ \left[\exp\left(-i(a+b)/2\right) \ \Psi\right]}}\right)}
+{\hbar}\hskip 2pt{{b-a}\over{2}}.
\end{equation}
The corresponding Bohm's relation can be easily obtained by 
taking in this last relation $|\alpha|=1$ and $|\beta|=a=b=0$ 
($\alpha=1, \beta=0$). Since the stationary SE
\begin{equation}
-{\hbar^2\over2m}{\Delta\psi}+V\left(x,y,z\right)\psi=E\psi,
\end{equation}
is linear, and taking into account the fact that for any 
solution $\phi$ of relation (10) then 
${\rm \ Re}\hskip 2pt\phi$ and ${\rm \ Im} \hskip 2pt\phi$ 
are also solutions, expression (9) and its corresponding 
Bohm's one strongly suggest to search for the QSHJE 
(eqs.(1) and (2)) a solution in the following form
\begin{equation}
S_0=\hbar\hskip2pt \arctan{\left(\psi_1\over\psi_2\right)}
+\hbar l,
\end{equation}
where $\psi_1$ and $\psi_2$ are two real independent solutions 
of the  SE, eq. (10), and $l$ an arbitrary dimensionless 
constant. Setting
\begin{equation}
U={{\psi_{1}}\over{\psi_{2}}},
\end{equation}
we have
\begin{equation}
\vec\nabla{S_0}={{\hbar \ \vec\nabla{U}}\over{1+U^2}}.
\end{equation}
Substituting this expression in (2), we obtain
\begin{equation}
{2\vec\nabla{R}\cdot\vec\nabla{U}}-
{{2UR}\over{1+U^2}}\left(\vec\nabla U\right)^2+R\Delta U=0.
\end{equation}
Using the fact that $\psi_1$ and $\psi_2$ solve $(10)$, 
from $(12)$ we can deduce that 
\begin{equation}
{\Delta{U}}=-2{{\vec \nabla\psi_{2}}
\over{\psi_{2}}}\cdot\vec {\nabla }{U}.
\end{equation}
It follows that relation $(14)$ takes the form
\begin{equation}
{\left({{\vec \nabla R}\over{R}}-
{{U\vec \nabla U}\over{1+U^{2}}}-
{{\vec \nabla \psi_{2}}\over{\psi_{2}}}\right)}\cdot{\vec {\nabla } {U}}=0.
\end{equation}
As $\psi_1$ and $\psi_2$ are independent solutions of (10), 
in general $\vec \nabla U$ does not vanish and is not 
perpendicular to the vector in brackets appearing in $(16)$. 
It follows that
\begin{equation}
{{\vec \nabla R}\over R}={{U\vec \nabla U}\over{1+U^2}}+
{{\vec \nabla \psi_2}\over{\psi_2}}.
\end{equation}
Substituting this relation in the identity
\begin{equation}
{{\Delta R}\over R}={{\vec\nabla}\cdot\left({\vec \nabla R}
\over {R}\right)}+{\left({\vec \nabla R}\over R\right)^2}
\end{equation}
and using (15), we obtain
\begin{equation}
 {{\Delta R}\over R}= 
{\left({\vec\nabla U}\over{1+U^2}\right)^2}
+{{\Delta \psi_2}\over {\psi_2}}.
\end{equation}
Using (13) and taking into account the fact that 
$\psi_2$ solves (10), relation (19) becomes
\begin{equation}
 {{\Delta R}\over R}= {\left({\vec\nabla S_0}\over{\hbar}\right)^2}
+{{2m\left(V-E\right)}\over{\hbar^2}}.
\end{equation}
This expression is equivalent to eq. (1). This result means 
that expression (11) for $S_0$ is a solution of the QSJHE, 
(eqs. (1) and (2)).

Now, let us determine the expression of $R$. Substituting (12) in 
(17), we obtain 
\begin{equation}
{{\vec\nabla} R \over R}= {{1}\over{2}}{{\vec \nabla \left(
{\psi_1^2}+{\psi_2^2}\right)}\over{\left({\psi_1^2}+{\psi_2^2}\right)}},
\end{equation}
which leads to
\begin{equation}
{\vec\nabla}{\left[\ln{R\over{\sqrt{\psi_{1}^2+\psi_{2}^2}}}\right]}=\vec0.
\end{equation}
Finally, we obtain
\begin{equation}
R=c\sqrt{{\psi_1^2}+{\psi_2^2}} \ ,
\end{equation}
where $c$ is an integration constant.

Of course, a direct substitution of expressions (11) and 
(23) in (1) and (2) allows, with the use of the SE, 
to check that these expressions are indeed solutions 
of (1) and (2).

\section {The case of separated variables }

\subsection{The one-dimensional case}

Before we examine the higher-dimensional cases, it is instructive 
to consider the problem of identifying the integration constants 
in one dimension. The SE, eq. $(10)$, reduces to
\begin{equation}
-{\hbar ^2\over2m}{d^2\psi\over{dx^2}}+V\left(x\right)\psi=E\psi .
\end{equation}
Let $\left(\phi_{1},\phi_{2}\right)$ be a set of two real independent 
solutions of (24). Since the SE is linear and always admits two 
real independent solutions, in order to make  
visible all the integration constants in (11), let us write the 
real functions $\psi_1$ and $\psi_2$ in the general form
\begin{equation}
\psi_{1}=\nu_{1}\phi_{1}+\nu_{2}\phi_{2},
\hskip30pt \psi_{2}=\mu_{1}\phi_{1}+\mu_{2}\phi_{2}, 
\end{equation}
where $\left(\nu_{1},\nu_{2},\mu_{1},\mu_{2}\right)$ are arbitrary 
real constants satisfying the condition $\nu_{1}\mu_{2}$ $\not=$
$\nu_{2}\mu_{1}$ which guarantees that $\psi_1$ and $\psi_2$ are
independent. Relation (11) turns out to be
\begin{equation}
S_0=\hbar \hskip2pt \arctan\left({\nu_1\phi_1+\nu_2\phi_2}
\over{\mu_1\phi_1+\mu_2\phi_2}\right)+\hbar l.
\end{equation}
This expression must be a general solution of the one-dimensional 
QSHJE, eq. (6). As explained in ref. \cite{b1}, eq. (6) is a 
second order differential equation with respect to 
$\partial{S_0}/\partial{x}$, then this derivative must 
depend only on two integration constants. Therefore, the 
function $S_0$ contains a further constant which must be 
additive. It is represented by $\hbar l$ in (26). 
Thus, we can set $\nu_{1}=\mu_{2}=1$ and interpret 
$\left(\mu_{1},\nu_{2},\hbar l\right)$ as integration 
constants of $S_{0}$. 

However, the extension of this reasoning to higher dimensions 
is not trivial even in the case of separated variables. 
In fact, in two or three dimensions we have not an ordinary 
differential equation, but a couple of two partial differential 
equations, eqs. (1) and (2). For example in three dimensions, 
where the potential takes the following form
$
V(x,y,z) = V_{x}(x) + V_{y}(y) + V_{z}(z),
$
if we tempt to search for eqs. (1) and (2) solutions with the 
standard method by writing $S_0$ in the form
\[
S_0(x,y,z) = S_{0x}(x) + S_{0y}(y) + S_{0z}(z) ,
\]
and use for $R$ a form as the one given in \cite{bfm}
\[
R(x,y,z) = R_{x}(x)  R_{y}(y)  R_{z}(z),  
\]
the three separated equations which result from eqs. (1) and 
(2) differ from the usual one-dimensional QSHJE and lead to a 
deadlock. This is the reason for which we will 
resolve this problem with a novel approach. 
We first review the problem in one dimension and reproduce the 
expected results. This approach consists 
in determining the minimum number of parameters in the set 
$\left(\nu_1,\nu_2,\mu_1,\mu_2 \right)$, which we must keep 
free in the expression of $S_0$, but sufficient to guarantee 
the invariance of the conjugate momentum
\begin{equation}
{\partial{S_0}\over\partial{x}}={\partial{\tilde{S}_{0}}
\over\partial{x}}
\end{equation} 
under an arbitrary linear transformation of the couple 
$\left(\phi_{1},\phi_{2}\right)$
\begin{equation}
\phi_{i}\rightarrow\theta_{i}=\sum_{j=1}^{2}\alpha_{ij}\phi_{j},
\hskip20pt i=1,2
\end{equation} 
$\tilde{S}_{0}$ being the new reduced action and $\alpha_{ij}$ 
arbitrary real constant parameters. In other words, if we choose 
another couple $\left(\theta_{1},\theta_{2}\right)$ of the 
solutions of SE instead of $\left(\phi_{1},\phi_{2}\right)$ 
and write the reduced action in the same form as in (26),
\begin{equation}
{\tilde{S}_{0}}=\hbar\hskip 2pt \arctan \left({
\tilde{\nu}_{1}\theta_{1}+\tilde{\nu}_{2}\theta_{2}}
\over{\tilde{\mu}_{1}\theta_{1}+\tilde{\mu}_{2}\theta_{2}}
\right)+\hbar \tilde{l},
\end{equation} 
the equation of motion, as relation (7), must remain unchanged, 
meaning that our mathematical choice does not affect the 
physical result. In this procedure, 
we have to accomplish two principal simultaneous tasks. The first 
is to prove for any transformation (28) the existence of the 
parameters $\left(\tilde{\nu}_1,\tilde{\nu}_2,\tilde{\mu}_1,
\tilde{\mu}_2 \right)$, used in $\tilde{S}_0$ instead of 
$\left(\nu_1,\nu_2,\mu_1,\mu_2 \right)$, in such a way as 
to guarantee the invariance (27). The second task consists 
in eliminating the maximum of parameters in the set 
$\left(\nu_1,\nu_2,\mu_1,\mu_2 \right)$ as in 
$\left(\tilde{\nu}_1,\tilde{\nu}_2,\tilde{\mu}_1,
\tilde{\mu}_2 \right)$ without violating relation (27) 
and without inducing any restriction on transformation (28).
We would like to add that this idea of invariance was first 
introduced in ref. \cite{bd2}. However the goal in this 
reference was not to determine the minimum number of 
pertinent parameters which would be playing the role of 
integration constants, but it was only to check that the 
choice of the couple of solutions $(\phi_1,\phi_2)$ does 
not affect the equation of motion.

Before going further, it is crucial to note that the 
extension to higher dimensions of the  
invariance of the derivatives of $S_0$ will induce 
insurmountable calculations. This is the reason for which 
we turn condition (27) into the invariance of $S_0$,
\begin{equation}
S_0=\tilde{S}_{0}+\hbar l_0,
\end{equation} 
up to an additive constant $\hbar l_o$. It is 
then interesting to remark that for two arbitrary functions 
$f$ and $g$, if 
\begin{equation}
\arctan (f)=\arctan (g) + l_1,
\end{equation} 
we can easily deduce that
\begin{equation}
f-g=k_{1} \left(1+fg\right),
\end{equation} 
where $k_1 =\tan \left(l_1\right)$. Then, substituting (26) and 
(29) in (30) and applying (31) and (32), we deduce
\begin{equation}
\sum_{i=1}^{2}\sum_{j=1}^{2}\left[k\left(\mu_{i}\tilde{\mu}_{j}
+\nu_{i}\tilde{\nu}_{j}\right)+\tilde{\nu}_{j}\mu_{i}
-\nu_{i}\tilde{\mu}_{j}\right]\phi_{i}\theta_{j}=0,
\end{equation} 
where
\begin{equation}
k=\tan\left(l_{0}+\tilde{l}-l\right).
\end{equation}
From transformation (28), we can deduce $\phi_i$
\begin{equation}
\phi_{i}=\sum_{l=1}^{2}\beta_{il}\hskip 2pt \theta_{l},
\end{equation}
where $\beta_{ij}$ can be determined by the system of four equations
\begin{equation}
\sum_{j=1}^{2}\beta_{ij}\alpha_{jl}=\delta_{il}.
\end{equation}
Substituting (35) in (33), we find
\begin{equation}
\sum_{j=1}^{2}\sum_{l=1}^{2}
\left[\tilde{\mu}_{j}\sum_{i=1}^{2}\left(k\mu_{i}
-\nu_{i}\right)\beta_{il}+\tilde{\nu}_{j}\sum_{i=1}^{2}
\left(k\nu_{i}+\mu_{i}\right)\beta_{il}\right]\theta_{l}\theta_{j}=0.
\end{equation}
This equation contains three independent terms and then the 
coefficients which precede $\theta_{1}^{2}$, $\theta_{2}^{2}$ 
and $\theta_{1}\theta_{2}$ must take a vanishing value
\begin{equation}
\tilde{\mu}_{1}\sum_{i=1}^{2}\left(k\mu_{i}-\nu_{i}\right)
\beta_{i1}+\tilde{\nu}_{1}\sum_{i=1}^{2}\left(k\nu_{i}+\mu_{i}
\right)\beta_{i1}=0, \hskip25mm 
\end{equation}
\begin{equation}
\tilde{\mu}_{2}\sum_{i=1}^{2}\left(k\mu_{i}-\nu_{i}\right)
\beta_{i2}+\tilde{\nu}_{2}\sum_{i=1}^{2}\left(k\nu_{i}+\mu_{i}
\right)\beta_{i2}=0, \hskip25mm 
\end{equation}
\begin{eqnarray}
\tilde{\mu}_{1}\sum_{i=1}^{2}\left(k\mu_{i}-\nu_{i}\right)\beta_{i2}+
\tilde{\nu}_{1}\sum_{i=1}^{2}\left(k\nu_{i}+\mu_{i}\right)\beta_{i2}
\hskip30mm&& \nonumber\\
+\tilde{\mu}_{2}\sum_{i=1}^{2}\left(k\mu_{i}-\nu_{i}\right)\beta_{i1}
+\tilde{\nu}_{2}\sum_{i=1}^{2}\left(k\nu_{i}+\mu_{i}\right)\beta_{i1}=0.
\end{eqnarray}
Thus, we have three independent equations and four unknown parameters: 
$\tilde{\nu}_1$,  $\tilde{\nu}_2$, $\tilde{\mu}_1$ and
$\tilde{\mu}_2$. However, if we divide by $\tilde{\nu}_1$ 
in the quotient appearing in (29) and define 
$\tilde{\nu}_2/\tilde{\nu}_1$, $\tilde{\mu}_1/\tilde{\nu}_1$ and 
$\tilde{\mu}_2/\tilde{\nu}_1$ as new parameters, it amounts to setting 
$\tilde{\nu}_1=1$ and keeping $\tilde{\nu}_2$, $\tilde{\mu}_1$ and 
$\tilde{\mu}_2$ unchanged. Of course, we must also take $\nu_{1}=1$ 
since the same form for $S_0$ and $\tilde{S}_0$ is required.  
Furthermore, the parameter $k$ defined by (34) is free. 
A judicious choice of $k$ allows to fix one of the parameters 
$\tilde{\nu}_2$, $\tilde{\mu}_1$ and $\tilde{\mu}_2$. 
For example, if we take
\begin{equation}
{k}={{\sum_{i=1}^{2}\left(\nu_{i}\beta_{i1}
-\mu_{i}\beta_{i2}\right)}\over{\sum_{i=1}^{2}\left(\nu_{i}\beta_{i2}
+\mu_{i}\beta_{i1}\right)}},
\end{equation}
we can check that the system (38), (39) and (40) gives 
$\tilde\mu_{2}=1$ and allows to express $\tilde\nu_{2}$ and 
$\tilde\mu_{1}$ in terms of $\mu_{i}$, $\nu_{i}$ and $\beta_{ij}$. 
Of course, for the same reason as above, we must also take 
$\mu_{2}=1$. As $\beta_{ij}$ can be expressed in terms of 
$\alpha_{ij}$ from (36), if we add any condition on $\tilde\nu_{2}$ 
or $\tilde\mu_{1}$, the system (38), (39) and (40) will 
induce a relation between $\mu_{1}$, $\nu_{2}$ and $\alpha_{ij}$. 
Since $\mu_{1}$ and $\nu_{2}$ correspond to the initial choice 
$(\phi_1, \phi_2)$ used in the initial reduced 
action, this relation between 
$\mu_{1}$, $\nu_{2}$ and $\alpha_{ij}$ will represent a 
restriction on transformation (28) and then on 
the choice of the couple $(\theta_{1},\theta_{2})$. 
Thus, if we want to guarantee the invariance of 
$\partial{S_0}/\partial{x}$ under any linear transformation, 
we must keep free $(\nu_{2},\mu_{1})$ in the expression of 
$S_0$, as we must do it for $(\tilde{\nu}_{2},\tilde{\mu}_{1})$ 
if we choose to deal with $\tilde{S}_0$. Finally, we reach the same 
conclusion as at the beginning of this section where these two 
pertinent parameters $(\nu_2,\mu_1)$ are identified as integration  
constants of the reduced action $S_0$, expression (26).

\subsection{The two-dimensional case}

Let us consider the two-dimensional case in which the potential 
takes the form
\begin{equation}
V\left(x,y\right)=V_{x}\left(x\right)+V_{y}\left(y\right).
\end{equation}
Writing $\psi\left(x,y\right) =X\left(x\right)Y\left(y\right)$, 
the two-dimensional SE leads to 
\begin{equation}
-{\hbar^{2}\over2m}{d^{2}X\over{dx^{2}}}+V_{x}X=E_{x}X,
\end{equation}
\begin{equation}
-{\hbar^{2}\over2m}{d^{2}Y\over{dy^{2}}}+V_{y}Y=E_{y}Y,
\end{equation} 
where $E_x$ and $E_y$ are real constants satisfying
\begin{equation}
E_{x}+E_y=E.
\end{equation} 
Let us call $\left(X_{1},X_{2}\right)$ and $\left(Y_{1},
Y_{2}\right)$ two couples of real independent solutions 
respectively of (43) and (44). Then, the two-dimensional 
SE admits four independent solutions
\begin{equation}
\phi_{1}=X_{1}Y_{1},\hskip20pt \phi_{2}=X_{1}Y_{2},
\hskip20pt \phi_{3}=X_{2}Y_{1},\hskip20pt \phi_{4}=X_{2}Y_{2}.
\end{equation} 
As in one dimension, the general form of the reduced action 
is obtained from (11) by writing
\begin{equation}
S_0=\hbar \hskip 2pt \arctan \left({\sum_{i=1}^{4}\nu_i\phi_i}
\over{\sum_{i=1}^{4}\mu_i\phi_i}\right)+\hbar l,
\end{equation}
$\left(\nu_1,...,\nu_4,\mu_1,...,\mu_4\right)$ being arbitrary 
real constants satisfying a condition with which $\psi_1$ and 
$\psi_2$ are not proportional. The equations of motion, as relations 
(8), are now obtained from $\partial{S_0}/\partial{x}$ and 
$\partial{S_0}/\partial{y}$. Then, let us impose the invariance 
of the conjugate momentum components 
\begin{equation}
{\partial{S_0}\over\partial{x}}={\partial{\tilde{S}_0}
\over\partial{x}}\ ,\hskip30pt{\partial{S_0}\over\partial{y}}=
{\partial{\tilde{S}_0}\over\partial{y}}
\end{equation}
under the following arbitrary linear transformation
\begin{equation}
\phi_{i} \rightarrow \theta_{i}=\sum_{j=1}^{4}\alpha_{ij}\phi_{j},
\hskip20pt i=1,2,3,4
\end{equation}
where $\tilde{S_0}$ is the new reduced action defined 
as in (47)
\begin{equation}
\tilde{S}_0=\hbar \hskip 2pt \arctan \left({\sum_{i=1}^{4}
\tilde{\nu}_i\theta_i}\over{\sum_{i=1}^{4}\tilde{\mu}_i\theta_i}
\right)+\hbar \tilde{l},
\end{equation}
and $\alpha_{ij}$ are arbitrary real constant parameters.
It is easy to show that conditions (48) can be turned 
into $S_0=\tilde{S}_{0}+\hbar l_0$ as in (30). 
Thus, with the use of (31) and (32), we obtain
\begin{equation}
\sum_{i=1}^{4}\sum_{j=1}^{4}\left[k\left(\mu_{i}\tilde{\mu}_{j}
+\nu_{i}\tilde{\nu}_{j}\right)+\mu_{i}\tilde{\nu}_{j}
-\nu_{i}\tilde{\mu}_{j}\right]\phi_{i}\theta_{j}=0,
\end{equation}
where $k$ is given as in (34). In contrast to the one-dimensional 
case and for reasons that will be clarified farther, 
we will make from (51) an expansion in
$\phi_{i} \phi_{l}$ and  not in $\theta_{i} \theta_{l}$. 
Thus, substituting (49) in (51), we get to
\begin{equation}
\sum_{i=1}^{4}\sum_{l=1}^{4}\left[
\nu_{i}\sum_{j=1}^{4}\left(k\tilde{\nu}_{j}-
\tilde{\mu}_{j}\right)\alpha_{jl}+\mu_{i}\sum_{j=1}^{4}
\left(k\tilde{\mu}_{j}+\tilde{\nu}_{j}\right)\alpha_{jl}
\right]\phi_{i}\phi_{l}=0.
\end{equation}
Setting
\begin{equation}
A_{l}=\sum_{j=1}^{4}\left(k\tilde{\nu}_{j}-\tilde{\mu}_{j}
\right)\alpha_{jl},\hskip20pt B_{l}=\sum_{j=1}^{4}
\left(k\tilde{\mu}_{j}+\tilde{\nu}_{j}\right)\alpha_{jl},
\end{equation}
eq. (52) becomes
\begin{equation}
\sum_{i=1}^{4}\sum_{l=1}^{4}\left(A_{l}\hskip 2pt\nu_{i}
+B_{l}\hskip 2pt \mu_{i}\right)\phi_{i}\phi_{l}=0.
\end{equation}
In this equation, we have sixteen terms (4x4=16). The symmetry 
$\phi_{i}\phi_{l}$ $=$ $\phi_{l}\phi_{i}$ reduces the number of 
terms to ten $[(16-4)/2+4=10]$. From $(46)$, we have also 
$\phi_{1}\phi_{4}=\phi_{2}\phi_{3}$. Thus, in (54) 
we have nine independent terms. For $(i,l)\not\in\{(1,4),
(4,1),(2,3),(3,2)\}$, we have eight equations
\begin{equation}
A_{l}\hskip 2pt\nu_{i}+B_{l}\hskip 2pt\mu_{i}+A_{i}
\hskip 2pt\nu_{l}+B_{i}\hskip 2pt\mu_{l}=0.
\end{equation}
In the case where $i=l$, (55) gives four equations
\begin{equation}
{\mu_{i}}=-{A_{i}\over{B_{i}}}\nu_{i}, \hskip20pt i=1, 2 , 3, 4
\end{equation}
and then, by taking into account this result, (55) 
gives the other four equations for $i \not =l$
\begin{equation}
{\nu_{i}\over{B_{i}}}={\nu_{l}\over{B_{l}}},
\end{equation}
meaning that $\nu_{1}/B_{1}=\nu_{2}/B_{2}=\nu_{3}/B_{3}=
\nu_{4}/B_{4}$. As $\phi_{1}\phi_{4}=\phi_{2}\phi_{3}$, 
the ninth equation is obtained by combining
the cases $(i,l)=(1,4)$ and $(i,l)=(2,3)$
\begin{equation}
A_{4}\hskip2pt \nu_{1}+B_{4}\hskip2pt \mu_{1}+A_{1}
\hskip2pt \nu_{4}+B_{1}\hskip2pt \mu_{4}+A_{3}\hskip2pt \nu_{2}
+B_{3}\hskip2pt \mu_{2}+A_{2}\hskip2pt \nu_{3}+B_{2}\hskip2pt \mu_{3}=0.
\end{equation}
With the use of (56) and (57), it is easy to check that 
(58) represents an identity. In (56) we have four independent 
equations but in (57) only three. Thus, 
we have seven linear equations and eight unknown parameters: 
$(\tilde{\nu}_{1},...,\tilde{\nu}_{4},\tilde{\mu}_{1},...,
\tilde{\mu}_{4})$ which are present through $A_i$ and 
$B_i$. However, as in one dimension, the operation 
consisting in dividing by $\tilde{\nu}_{1}$ in the quotient 
appearing in (50) amounts to setting 
$\tilde{\nu}_{1}=1$ 
in (50) and $\nu_{1}=1$ in (47). Now, we have seven equations 
with seven unknown parameters. In addition, $k$ being free, we 
can choose its value in order to obtain for example 
$\tilde{\mu}_{4}=1$, and then to also take   
$\mu_{4}=1$. This amounts to solving our system of seven 
equations with respect to the following seven parameters:
$(\tilde{\nu}_{2},\tilde{\nu}_{3},\tilde{\nu}_{4},
\tilde{\mu}_{1},\tilde{\mu}_{2},\tilde{\mu}_{3}, k)$.
As in one dimension, any further condition on 
$(\tilde{\nu}_{2},\tilde{\nu}_{3},\tilde{\nu}_{4},
\tilde{\mu}_{1},\tilde{\mu}_{2},\tilde{\mu}_{3})$ 
will induce restrictions on the linear transformation (49). 
In conclusion, the number of pertinent parameters 
of the reduced action (47), with which we can reproduce the 
same equations of motion 
under any linear transformation, is six. As we have fixed 
$\nu_{1}=\mu_{4}=1$, the pertinent parameters playing the role 
of integration constants are $\left(\nu_{2},\nu_{3},\nu_{4},
\mu_{1},\mu_{2},\mu_{3}\right)$. 

We would like to add that, in contrast to the one-dimensional 
case, we have made in (52) an expansion in  $\phi_{i} \phi_{l}$ 
but not in $\theta_{i} \theta_{l}$. The reason is that in two dimensions, 
the ten product $\phi_{i}\phi_{l}$ are not linearly independent since 
we have seen that $\phi_{1}\phi_{4}=\phi_{2}\phi_{3}$. This implies 
a more complicated relation between the ten products 
$\theta_{i} \theta_{l}$ and will induce a tedious calculations 
if we make the expansion in $\theta_{i}\theta_{l}$.

\subsection{The three-dimensional case}

In the same manner, let us now consider the three-dimensional case 
and write the potential in the following form
\begin{equation}
V(x,y,z) = V_{x}(x) + V_{y}(y)
+ V_{z}(z).
\end{equation}
Writing $ \psi (x,y,z) = X(x)Y(y)Z(z) $, 
the SE in three dimensions, eq. (10), leads to 
\begin{equation}
-{\hbar^{2} \over 2m}{d^{2}X \over {dx^{2}}}+V_{x}X=E_{x}X,
\end{equation} 
\begin{equation}
-{ \hbar^{2} \over 2m}{d^{2}Y \over {dy^{2}}}+V_{y}Y=E_{y}Y,
\end{equation} 
\begin{equation}
-{\hbar^{2} \over 2m}{d^{2}Z \over {dz^{2}}}+V_{z}Z = E_{z}Z,
\end{equation} 
where $E_{x}$, $E_{y}$ and $E_{z}$ are real constants satisfying 
\begin{equation}
E_{x}+E_{y}+E_{z}=E.
\end{equation} 
Let us call 
    $ \left( X_{1},X_{2} \right) $, 
    $ \left( Y_{1},Y_{2} \right) $ 
and 
    $ \left( Z_{1},Z_{2} \right) $ 
three couples of real independent solutions respectively of 
(60), (61) and (62). It follows that the SE in three dimensions  
admits eight real independent solutions
\begin{eqnarray}
\left\{ \begin{array}{cc}
               \phi_{1}=X_{1}Y_{1}Z_{1}, \hskip6pt
               \phi_{2}=X_{1}Y_{1}Z_{2}, \hskip6pt
               \phi_{3}=X_{1}Y_{2}Z_{1}, \hskip6pt
               \phi_{4}=X_{1}Y_{2}Z_{2},    \\ 
               \phi_{5}=X_{2}Y_{1}Z_{1}, \hskip6pt
               \phi_{6}=X_{2}Y_{1}Z_{2}, \hskip6pt
               \phi_{7}=X_{2}Y_{2}Z_{1}, \hskip6pt
               \phi_{8}=X_{2}Y_{2}Z_{2}.  
           \end{array}  \right.
\end{eqnarray}
The general form of the reduced action can be deduced from (11) 
by writing 
\begin{equation}
S_0= \hbar \hskip 2pt \arctan \left( {{\sum_{i=1}^{8} \nu_i \phi_i}
 \over {\sum_{i=1}^{8}{\mu}_i\phi_i} } \right) + \hbar l.
\end{equation}
As in the previous cases, in order to determine the pertinent 
parameters among the real constants $(\nu_{1},...,\nu_{8},
\mu_{1},...,\mu_{8})$, let us impose the invariance of the 
conjugate momentum components
\begin{equation}
{\partial {S_0} \over \partial{x}}=
          {\partial {\tilde{S}_0} \over \partial {x}},\hskip20pt
{\partial {S_0} \over \partial{y}}=
          {\partial {\tilde{S}_0} \over \partial {y}},\hskip20pt
{\partial {S_0} \over \partial{z}}=
          {\partial {\tilde{S}_0} \over \partial {z}}
\end{equation}
under the following arbitrary linear transformation
\begin{equation}
\phi_{i} \rightarrow \theta_{i}=\sum_{j=1}^{8}
\alpha_{ij}\phi_{j}, \hskip20pt i=1,...,8
\end{equation} 
where $\tilde{S}_0$ is the new reduced action
defined as in (65)
\begin{equation}
\tilde{S}_0 = \hbar \hskip 2pt \arctan \left({
{\sum_{i=1}^{8}
    \tilde{\nu}_i\theta_i} \over 
    {\sum_{i=1}^{8} \tilde {\mu}_{i} \theta_{i}} 
                                   } \right) + \hbar  \tilde{l} ,
\end{equation}
and $\alpha_{ij}$ are arbitrary real constant parameters.
Conditions (66) can be also turned into 
$S_0= \tilde {S}_0 + \hbar l_0$. Thus, as in two 
dimensions, we deduce that
\begin{equation}
\sum_{i=1}^{8}\sum_{l=1}^{8} (A_{l}\hskip 2pt \nu_{i}
+B_{l} \hskip 2pt \mu_{i} ) \phi_{i}\phi_{l}=0,
\end{equation}
where
\begin{equation}
A_{l}=\sum_{j=1}^{8} (k \tilde{\nu}_{j}
- \tilde{\mu}_{j} )\alpha_{jl},\hskip20pt B_{l}
=\sum_{j=1}^{8} (k \tilde{\mu}_{j} + \tilde{\nu}_{j} ) \alpha_{jl}.
\end{equation}
In (69), we have sixty-four terms (8x8=64). The symmetry 
$\phi_{i}\phi_{l}=\phi_{l}\phi_{i}$ reduces this number to 
thirty-six $[(64-8)/2+8=36]$. From (64), all possible relations 
which we can deduce between the products $\phi_{i}\phi_{l}$ are
\begin{eqnarray}
\left\{ \begin{array}{cc}
               \phi_{1}\phi_{4}=\phi_{2}\phi_{3},\hskip10pt
               \phi_{1}\phi_{6}=\phi_{2}\phi_{5},\hskip10pt
               \phi_{1}\phi_{7}=\phi_{3}\phi_{5},    \\ 
               \phi_{2}\phi_{8}=\phi_{4}\phi_{6},\hskip10pt
               \phi_{3}\phi_{8}=\phi_{4}\phi_{7},\hskip10pt
               \phi_{5}\phi_{8}=\phi_{6}\phi_{7} 
           \end{array}  \right.
\end{eqnarray}
and
\begin{equation}
\phi_{1}\phi_{8}=\phi_{2}\phi_{7}=\phi_{3}\phi_{6}=\phi_{4}\phi_{5}.
\end{equation} 
In (71) and (72) we have nine independent relations. It follows 
that in (69), we have twenty-seven independent terms $(36-9=27)$. 
For the terms $i=l$, we deduce eight relations
\begin{equation}
A_{i}\hskip 2pt\nu_{i}+B_{i}\hskip 2pt\mu_{i}=0,
\end{equation}
leading to
\begin{equation}
{\mu_{i}}=-{A_{i} \over {B_{i}}} \nu_{i},  \hskip12pti=1,...,8.
\end{equation}
If
\begin{eqnarray}
(i,l) \ \in \ \left\{ (1,2),\ (1,3),\ (1,5),\ (2,4),\ (2,6),
\ (3,4),\right. \hskip20mm&&  \nonumber\\
\left. (3,7),\ (4,8),\ (5,6),\ (5,7),\ (6,8),\ (7,8) \right\},
\end{eqnarray}
from (69) we deduce the following twelve relations
\begin{equation}
A_{l}\hskip 2pt\nu_{i}+B_{l}\hskip 2pt\mu_{i}+A_{i}
\hskip 2pt\nu_{l}+B_{i}\hskip 2pt\mu_{l}=0.
\end{equation}
Taking into account relations (74), these last relations lead to 
\begin{equation}
{\nu_{i} \over {B_{i}}}={ \nu_{l} \over {B_{l}}}.
\end{equation}
If we look more closely at the set given in (75), we deduce 
that (77) is valid $\forall \hskip3pt i \in \left[1,2,...,8\right]$ 
and $\forall \hskip3pt l \in\left[1,2,...,8\right]$.
By using the couples of indexes appearing in (71) and (72), 
we can deduce the seven remaining relations $(27-8-12=7)$. 
We can check that they all represent identities. For example, 
with the first relation in (71), we deduce from (69)
\begin{equation}
A_{4}\hskip 2pt\nu_{1}+B_{4}\hskip 2pt\mu_{1}
+A_{1}\hskip 2pt\nu_{4}+
B_{1}\hskip 2pt\mu_{4}+A_{3}\hskip 2pt\nu_{2}+
B_{3}\hskip 2pt\mu_{2}+A_{2}\hskip 2pt\nu_{3}+B_{2}\mu_{3}=0.
\end{equation}
If we substitute in this relation $\mu_{1},\mu_{2},\mu_{3}$ and 
$\mu_{4}$ by their values given in (74) and take into account (77), 
we easily obtain an identity. We emphasize that with (72), 
we have only one relation which is also an identity.

The twelve relations (76) are reduced, with the use of (74), 
to seven independent equations given in (77). The five missing 
relations are identities. This means that the system (74) and 
(76) is turned into the system (74) and (77) which contains  
fifteen (8+7=15) independent linear equations and sixteen 
unknown parameters $\left(\tilde{\nu}_{1},...,
\tilde{\nu}_{8},\tilde{\mu}_{1},...,\tilde{\mu}_{8}\right)$ 
which are present through $A_i$ and $B_i$. 
However, as in the previous cases, the operation consisting in dividing 
by $\tilde{\nu}_{1}$ in the quotient appearing in (68) amounts 
to setting $\tilde{\nu}_{1}=1$ in $(68)$ and then $\nu_{1}=1$ 
in (65), since the same form for $S_0$ and $\tilde{S_0}$ is 
required. Again, the freedom in choosing $k$ allows 
us to fix another parameter. Then, if we choose for example 
$\tilde{\mu}_8=1$, we have also to take $\mu_8=1$. Of course, 
any further condition on $\tilde{\nu}_i$ or $\tilde{\mu}_i$ 
will induce restriction on the linear transformation (67). In 
conclusion, the number of pertinent parameters in (65) 
is fourteen, and with the above choice, these parameters 
are $\left(\nu_2,...,\nu_8,\mu_1,...,\mu_7 \right) $. They 
play a role of integration constants of the reduced action, 
expression (65). In addition, we notice that the same reasons 
as in two dimensions have not allowed us to make in (69) 
an expansion in $\theta_i\theta_l$.

\section{ Microstates}

In this section, we will examine if the initial conditions, 
which determine completely the Schr\"odinger wave function, 
are sufficient to determine all the pertinent parameters of 
the reduced action. In other words, knowing that the 
reduced action is the generator of motion, 
one may wonder if, to these initial conditions of the wave 
function, correspond one or many trajectories of 
the particle.

For this purpose, remark that the constant $c$ appearing in 
(23) can be absorbed in the parameters $\alpha$ and $\beta$ 
when we use relation (3) or (5). Thus, in any dimension, let us write
\begin{equation}
R=\sqrt{\psi_{1}^{2}+\psi_{2}^{2}}
\end{equation}
and substitute in (5) the reduced action $S_0$ by its 
expression (11)
\begin{eqnarray}
\Psi=R\left\{\left(|\alpha|+
|\beta|\right)\cos\left[{
\arctan\left({\psi_{1}}\over{\psi_{2}}\right)+l}
+{{a-b}\over{2}}\right] \right. \hskip20mm&& \nonumber\\
+\left. i\left(|\alpha|-
|\beta|\right)\sin\left[{\arctan\left({\psi_{1}}
\over{\psi_{2}}\right)+l}+{{a-b}\over{2}}\right]\right\},
\end{eqnarray}
where we have discarded the unimportant constant phase factor 
$\exp\left[i(a+b)/2\right]$. Since the additive constant 
$l$ appearing in $(11)$ has no dynamical effect, we can choose 
it equal to $(b-a)/2$. Therefore, with the use of (79) and 
some trigonometric relations, we have
\begin{equation}
R\hskip2pt \cos\left[\arctan\left({\psi_{1}}
\over{\psi_{2}}\right)\right]=\psi_{2},\hskip20pt 
R\hskip2pt \sin\left[\arctan\left({\psi_{1}}
\over{\psi_{2}}\right)\right]=\psi_{1}
\end{equation}
and eq. (80) turns out to be
\begin{equation}
\Psi=\left(|\alpha|+|\beta|\right)\psi_{2}
+i\left(|\alpha|-|\beta|\right)\psi_{1}.
\end{equation}
This relation is valid in any dimension. First, it allows to 
reproduce the well-known results in one dimension obtained in 
\cite{b1}. In fact, comparing (11) and (26), we have
\begin{equation}
\psi_{1}=\phi_{1}+\nu_{2}\phi_{2},\hskip20pt
\psi_{2}=\mu_{1}\phi_{1}+\phi_{2},
\end{equation}
where we have set $\nu_{1}=\mu_{2}=1$ as indicated in the 
previous section.
Substituting $(83)$ in $(82)$, we obtain
\begin{equation}
\Psi=\left[\mu_{1}\left(|\alpha|+|\beta|\right)
+i\left(|\alpha|-|\beta|\right)\right]\phi_{1}
+\left[\left(|\alpha|+|\beta|\right)
+i\nu_{2}\left(|\alpha|-|\beta|\right)\right]\phi_{2}.
\end{equation}
On the other hand, the wave function can be written as a 
linear combination of the two real independent solutions 
$\phi_{1}$ and $\phi_{2}$ 
\begin{equation}
\Psi=c_{1}\hskip 2pt\phi_{1}+c_{2}\hskip 2pt\phi_{2},
\end{equation}
where $c_{1}$ and $c_{2}$ are generally complex constants 
determined by the initial or boundary conditions of the wave 
function. Identification of (84) and (85) leads to 
\begin{equation}
c_{1}=\mu_{1}\left(|\alpha|+|\beta|\right)+i\left(|\alpha|-|\beta|\right),
\end{equation}
\begin{equation}
c_{2}=|\alpha|+|\beta|+i\nu_{2}\left(|\alpha|-|\beta|\right). \hskip10pt
\end{equation}
In the case where $|\alpha|\not=|\beta|$, separating the real part 
from the imaginary one in (86) and (87), we obtain a system of 
four equations which can be solved with respect to $|\alpha|$, 
$|\beta|$, $\mu_{1}$ and $\nu_{2}$. It follows that in the complex wave 
function case $\left(|\alpha|\not=|\beta|\right)$, the initial 
conditions of the wave function $\Psi$ fix univocally the reduced 
action. There is no trace of microstates. In the real wave function 
case $\left(|\alpha|=|\beta|\right)$, up to a constant phase factor, 
(86) and (87) do not allow to determine $\nu_{2}$. Thus, for a 
given physical state $\Psi$, we have a family of trajectories, 
specified by the different values of $\nu_{2}$, corresponding 
to microstates not detected by the SE. The same conclusion is also 
reached in ref. \cite{fl1a}. We would like to add that if we use 
the Bohm ansatz $(\alpha=1,\beta=0)$,  eqs. $(86)$ and $(87)$ imply 
that ${\rm \ Im}\hskip2pt c_{1}=1$ and ${\rm \ Re}\hskip2pt c_{2}=1$. 
This is an unsatisfactory result since $c_{1}$ and $c_{2}$ are 
fixed by the initial conditions of the wave function and, then, we 
must not obtain fixed values for ${\rm \ Im}\hskip2pt c_{1}$ and 
${\rm \ Re}\hskip2pt c_{2}$. This is also the proof that the presence 
of $\alpha$ and $\beta$  in the relation between the reduced 
action and the Schr\"odinger wave function, eq. (3), is 
necessary. However, among the four real parameters which define 
the complex number $\alpha$ and $\beta$, there are only two 
which are linked to the initial conditions of the wave function. 
The two others are superfluous. This has been seen in ref. 
\cite{b1} by showing that the functions $R$ and $S_0$ are invariant 
under a dilatation and a rotation in the complex space of  
expression (3) of the wave function. This invariance allowed to 
make a transformation which fixed the two superfluous degrees of 
freedom. In our above reasoning we have eliminated these 
superfluous parameters first by discarding in (80) the phase 
factor $\exp\left[i(a+b)/2\right]$ and second by choosing 
$(b-a)/2$ equal to the additive integration constant $l$ in (82). 
This means that we have fixed the phases $a$ and $b$ of 
$\alpha$ and $\beta$ and kept free $|\alpha|$ and $|\beta|$. 
These modulus are determined by the four real equations 
which can be deduced from (86) and (87), meaning that 
$|\alpha|$ and $|\beta|$ are linked to the initial conditions 
of the wave function.

The two-dimensional case is similar to the three-dimensional one. 
For this reason, we straightforwardly investigate microstates 
in three dimensions. Comparing (11) and (65), we have 
\begin{equation}
\psi_{1}=\phi_{1}+\sum_{i=2}^{8}\nu_{i}\hskip2pt\phi_{i},
\hskip10pt \psi_{2}=\sum_{i=1}^{7}\mu_{i}\hskip2pt\phi_{i}+\phi_{8},
\end{equation}
where we have set $\nu_{1}=\mu_{8}=1$ as indicated in section 3. 
The functions $\phi_{i}$ are defined  in $(64)$. Substituting (
88) in (82), we obtain
\begin{eqnarray}
\Psi=\left[\mu_{1}\left(|\alpha|+|\beta|\right)+i\left(|\alpha|
-|\beta|\right)\right]\phi_{1} 
\hskip45mm&& \nonumber\\
+\sum_{i=2}^{7}\left[\mu_{i}\left(|\alpha|+|\beta|\right)
+i\nu_{i}\left(|\alpha|-|\beta|\right)\right]\phi_{i}
\hskip15mm&& \nonumber\\
+\left[|\alpha|+|\beta|+i\nu_{8}\left(|\alpha|-
|\beta|\right)\right]\phi_{8}.
\end{eqnarray} 
As above, the wave function can be written as a 
linear combination of $\phi_{i}$ ($i=1,...,8$) 
\begin{equation}
\Psi=\sum_{i=1}^{8}c_{i} \phi_{i},
\end{equation}  
where $c_{i}$ are complex constants which can be determined by the 
boundary conditions of the wave function. Identification of (89) 
and (90) leads to 
\begin{equation}
c_{1}=\mu_{1}\hskip2pt\left(|\alpha|+|\beta|\right)
+i\left(|\alpha|-|\beta|\right), \hskip78pt
\end{equation}
\begin{equation}
c_{i}=\mu_{i}\hskip2pt\left(|\alpha|+|\beta|\right)
+i\nu_{i}\hskip2pt\left(|\alpha|-|\beta|\right), \hskip 15pt i=2,3,...,7
\end{equation}
\begin{equation}
c_{8}=|\alpha|+|\beta|+i\nu_{8}\hskip2pt\left(|\alpha|
-|\beta|\right). \hskip88pt
\end{equation}   
In the complex wave function case 
$\left(|\alpha|\not=|\beta|\right)$, separating the real part 
from the imaginary one, we obtain from these relations a system 
of sixteen equations which can be solved with respect to the 
sixteen following unknown ($|\alpha|$, $|\beta|$, $\nu_{2}$, $...$, 
$\nu_{8}$, $\mu_{1}$, $...$, $\mu_{7}$). As in one dimension, the 
knowledge of $\left(c_{1},...,c_{8}\right)$ is sufficient 
to fix univocally the reduced action. There is  no trace of
microstates. In the real wave function case 
$\left(|\alpha|=|\beta|\right)$, 
up to a constant phase factor, it is clear that the system (91), 
(92) and (93) does not allow to determine ($\nu_{2},\nu_{3},...,\nu_{8}$). 
As in one dimension, for a given physical state $\Psi$, we have a 
family of trajectories, specified by the values of ($\nu_{2},\nu_{3},
...,\nu_{8}$), corresponding to microstates not detected by the 
SE. We also reach the same conclusion in two-dimensions, namely 
microstates appear only in the real wave function case. We would 
like to add that also in higher dimensions, $|\alpha|$ and $|\beta|$ 
are linked to the initial conditions of the Schr\"odinger wave 
function since these modulus are fixed by the system (91), (92) 
and (93), or its analogous in two dimensions. As in one dimension, 
the phases $a$ and $b$ are superfluous and have been eliminated in 
(82).
 
\section{ Conclusion}

This work can be summarized in three main results.

The first concerns the resolution in three dimensions of the QSHJE 
represented by the couple of relations (1) and (2). For any 
external potential, the expressions of the couple of functions 
$\left(R,S_0\right)$ are given in terms of a couple of real 
independent solutions of the SE. In other words, we reduced 
a problem of two non-linear partial differential equations to one 
linear equation.

The second result concerns the identification in the case of 
separated variables of the pertinent parameters playing a role of 
integration constants of the reduced action. Of course, in one 
dimension, the solution of this problem is already known 
\cite{fm1b,b1,fm2a,fm2b,fl3}. However, in higher dimensions, we had to 
solve two coupled partial differential equations and, then, we had 
no means to fix the number of integration constants since 
the standard method does not work. We surmounted this 
difficulty by imposing the invariance of the reduced action, up 
to an additive constant, under an arbitrary linear 
transformation of the set of solutions of the SE. This amounts 
to requiring that, for any choice of the set of solutions of 
the SE, we reproduce the same equations of motion. In this 
procedure, our task consisted in determining the minimum 
number of parameters which we must keep free in such a way as to 
impose this invariance of the reduced action. We first applied 
this new procedure in one dimension and reproduced the expected 
results \cite{b1}. We then extended the approach to two and 
three dimensions.

The third principal result we obtained concerns microstates. We 
showed that in one dimension, as in higher dimensions, 
microstates appear only in the case where the  Schr\"odinger 
wave function is real, up to a constant phase factor. In higher 
dimensions, it is the first time that microstates are 
analytically described. 
As indicated in \cite{landau}, in the case where there is no 
degeneracy, bound states are described by real wave functions. 
Thus, bound states reveal microstates not detected by the  
Schr\"odinger wave function. As concluded in the one-dimensional case 
\cite{fl1a}, in higher dimensions the  Schr\"odinger wave function  
does not describe exhaustively quantum phenomena. The QSHJE is more 
fundamental.

\vskip\baselineskip
\noindent
{\bf Acknowledgments:}
The authors would like to thank E. R Floyd for valuable discussions 
and suggestions.

\bigskip
\bigskip

\noindent
{\bf REFERENCES}

\end {document}